\documentclass[aps,onecolumn]{revtex4-2}

\usepackage[T1]{fontenc}
\usepackage{utopia}

\usepackage{amsmath,amssymb,amsfonts,graphicx,tabularx,xcolor}

\usepackage[unicode=true,colorlinks=true]{hyperref}

\hypersetup{linkcolor=blue,citecolor=blue,urlcolor=blue}

\begin{document}

\title{Two-component parton fractional quantum Hall state in graphene}

\author{Ying-Hai Wu}
\email{yinghaiwu88@hust.edu.cn}
\affiliation{School of Physics and Wuhan National High Magnetic Field Center, Huazhong University of Science and Technology, Wuhan 430074, China}

\begin{abstract}
We study the $\nu={\pm}1/2$ fractional quantum Hall states in graphene observed by Zibrov et al. [Nat. Phys. 14, 930 (2018)]. The parton construction is employed to provide a valley unpolarized trial wave function for these states. The lattice scale valley anisotropic terms in the Hamiltonian soften the repulsion between electrons in different valleys to favor the valley unpolarized state. The validity of our proposal is corroborated by numerical calculations.
\end{abstract}

\maketitle

\section{Introduction}

The extraordinary properties of graphene and other two-dimensional materials have sparked intensive research activities on these systems. One fruitful line of investigation is about strongly correlated many-body states in high magnetic field~\cite{Novoselov2005,ZhangYB2005}, for which the fractional quantum Hall (FQH) effect is a prominent example~\cite{Tsui1982}. The experimental signature of FQH effect in electric transport is quantized Hall resistance accompanied by exponentially suppressed longitudinal resistance. The application of an external magnetic field generates discrete Landau levels (LLs) for the electrons, where all single-particle orbitals in the same level have the same kinetic energy. If one Landau level is partially populated and electron-electron interactions are absent, there is an exponential degeneracy in the many-body spectrum. It is a great surprise that interactions would lead to energy gaps at certain filling factors $\nu$ (i.e., the number of electrons divided by the number of orbitals in each LL)~\cite{Laughlin1983,Haldane1983-3,Halperin1983,Jain1989-1,Moore1991}. 

The most extensively used platform for studying FQH states has been two-dimensional electron gases (2DEG) in GaAs based structures. A diverse zoo of FQH states have also been revealed in graphene and its multilayers thanks to substantial progress in sample fabrication~\cite{DuX2009,Bolotin2009,Dean2011,Feldman2012,KiDK2014,Maher2014,Kou2014,KimYW2015,Diankov2016,Zibrov2017,LiJIA2017,Zibrov2018,KimYW2019,LiuXM2019,LiJIA2019,HuangK2021}. The characteristics of graphene that distinguish it from conventional 2DEG include its Dirac dispersion, the four-fold spin and valley degeneracy, and direct access to the two-dimensional plane by experimental probes. The Dirac dispersion and valley degree of freedom endow the single-particle wave functions in the LLs with a spinor structure, which leads to effective interaction that could favor exotic FQH states with non-Abelian anyons~\cite{KimYW2019}. The long-range Coulomb interaction is SU(4) symmetric in the spin-valley space, but one generally expects spontaneous symmetry breaking such that the many-body states are spin and/or valley polarized. In addition, there are are lattice scale interactions that break the SU(4) spin-valley symmetry. The interplay of these effects has been analyzed to explain why some FQH states are missing or very weak in certain experiments~\cite{Feldman2012,Abanin2013,Sodemann2014}. 

This paper focuses on the $\nu={\pm}1/2$ FQH states in the zeroth LL of graphene~\cite{Zibrov2018}. To elucidate its nature, the first question is whether it is one-component or multi-component. In the former case, an odd denominator rule has been firmly established, which can be explained very well using the composite fermion theory~\cite{Jain1989-1}. An exception to this rule is the $5/2$ state in the second LL of GaAs~\cite{Willett1987}, which could harbor non-Abelian Ising anyons described by the Moore-Read state or its variants~\cite{Moore1991,Morf1998,Rezayi2000,Levin2007,LeeSS2007,Peterson2008-1,Peterson2008-2,Feiguin2009,Wojs2010,Storni2010,Pakrouski2015,Zaletel2015-1,Zucker2016,Balram2018-1,YangJ2017,Rezayi2017,Mishmash2018,Antonic2018,Simon2020,Rezayi2021}, but it is not likely to be relevant in the zeroth LL of graphene because the projected interaction here is quite different from that in the second LL of GaAs. For bilayer systems made of GaAs or graphene, even-denominator FQH states have been observed in many experiments~\cite{SuenYW1992,Eisenstein1992,LiuXM2019,LiJIA2019}, which can be understood using two-component Halperin and Jain wave functions~\cite{Halperin1983,HeS1993,Scarola2001,Peterson2010,Papic2010-1}. As we shall argue below, the Halperin state is not really a plausible candidate for the states observed in monolayer graphene. 

Instead, we propose that the experimental observation can be understood using the parton theory. The basic idea is to break one electron into multiple fermionic partons and each of them form an integer quantum Hall (IQH) state (or some other mean-field states in general)~\cite{Jain1989-2}. The partons are not completely free and should be glued together by emergent gauge fields~\cite{WenXG1991-1}. In spite of the elegance and generality of this approach, it does not seem to be relevant to actual quantum Hall physics for a long time. The composite fermion theory can be viewed as a special case of the parton construction, but this reinterpretation does not bring out particularly useful insights. In recent years, several works have found that some one-component parton FQH states might be realized in bilayer graphene~\cite{WuYH2017-1}, graphene~\cite{KimYW2019}, and second LL or wide quantum wells of GaAs~\cite{Balram2018-2,Faugno2019,Balram2020-1,Faugno2020-2,Balram2021}. In contrast, this paper aims to show that a two-component parton FQH state was realized in Ref.~\cite{Zibrov2018}. This state was originally constructed in Ref.~\cite{Jain1989-2} and tested as a candidate for the $5/2$ state in GaAs~\cite{Belkhir1993-1,Belkhir1993-2}. At present, most experimental and numerical results on the $5/2$ state suggest that it is actually spin-polarized~\cite{Morf1998,Feiguin2009,Stern2010,Stern2012,Tiemann2012,WangP2020}, so the two-component state is not likely relevant. It was proposed that this state features $d$-wave pairing~\cite{Moran2012}. We also note that an explanation completely different from ours has been proposed in Ref.~\cite{Narayanan2018}. 

The rest of this paper is organized as follows. In Sec.~\ref{model}, we introduce the model for graphene with valley anisotropic terms in the Hamiltonian. In Sec.~\ref{result}, we describe the parton FQH state and demonstrate that it could be realized in certain parameter regimes. The paper is concluded in Sec.~\ref{conclude}. 

\section{Models}
\label{model}

We consider a graphene sheet in the $x$-$y$ plane with the two Dirac cones denoted as $K^{+}$ and $K^{-}$. An external magnetic field is applied along the $z$ direction so the single-particle Hamiltonians in the two valleys are 
\begin{eqnarray}
H^{\pm}_{0} &=& v_{\rm F}
\left[
\begin{array}{cc}
0 & \pi_{x}{\pm}i\pi_{y} \\
\pi_{x}{\mp}i\pi_{y} & 0 
\end{array} 
\right],
\end{eqnarray}
where $v_{\rm F}$ is the Fermi velocity and $\pi_{x,y}$ are the canonical momentum operators. The zeroth LL of graphene includes the zero-energy single-particle eigenstates
\begin{eqnarray}
\left[
\begin{array}{c}
\phi_{m} \\
0 
\end{array} 
\right]
\end{eqnarray}
in the $K^{+}$ valley and
\begin{eqnarray}
\left[
\begin{array}{c}
0 \\
\phi_{m}
\end{array} 
\right]
\end{eqnarray}
in the $K^{-}$ valley, where $\phi_{m}$ represent the wave functions in the lowest LL of non-relativistic particles. It is customary to consider an infinite plane with electromagnetic vector potential $A=(-By/2,Bx/2,0)$. In this case, the argument of $\phi_{m}$ is the complex coordinate $z=x+iy$ and $m$ labels its angular momentum. Numerical calculations are commonly performed on the sphere rather than the disk because there is no complication due to open boundary~\cite{Haldane1983-3}. A radial magnetic field through the surface of the sphere is generated by a magnetic monopole at the center. The single-particle wave functions are labeled by the total angular momentum and its $z$-component~\cite{WuTT1976,WuTT1977}.

The zeroth LL is of our interest and all operators in the subsequent discussions are projected to it. The kinetic energy is a constant that may be neglected. The interaction Hamiltonian can be separated to the long-range Coulomb potential
\begin{eqnarray}
V(\mathbf{r}-\mathbf{r}') = \frac{1}{|\mathbf{r}-\mathbf{r}'|}
\end{eqnarray}
that respects the SU(4) spin-valley symmetry and some other terms that break this symmetry~\cite{Abanin2013}. The spin rotation symmetry is broken by the Zeeman coupling and the valley symmetry is broken by the following two terms
\begin{eqnarray}
&& H^{z}_{\rm VA} = \frac{g_{z}}{2} \int d\mathbf{r} : \left[ \widehat{\Psi}^{\dag}(\mathbf{r}) \tau^{z} \widehat{\Psi}(\mathbf{r}) \right]^{2} : \nonumber \\
&& H^{\perp}_{\rm VA} = \frac{g_{\perp}}{2} \int d\mathbf{r} \sum_{\alpha=x,y} : \left[ \widehat{\Psi}^{\dag}(\mathbf{r}) \tau^{\alpha} \widehat{\Psi}(\mathbf{r}) \right]^{2} :,
\end{eqnarray}
where the electron creation operator $\widehat{\Psi}^{\dag}(\mathbf{r})$ is a four-component spinor 
\begin{eqnarray}
\left[ \widehat{\psi}^{\dag}_{K^{+}\uparrow}(\mathbf{r}),\widehat{\psi}^{\dag}_{K^{+}\downarrow}(\mathbf{r}),\widehat{\psi}^{\dag}_{K^{-}\uparrow}(\mathbf{r}),\widehat{\psi}^{\dag}_{K^{-}\downarrow}(\mathbf{r}) \right].
\end{eqnarray}
$H^{z}_{\rm VA}$ arises from electron-electron interactions and $H^{\perp}_{\rm VA}$ arises from both electron-electron and electron-phonon interactions~\cite{JungJ2009,HouCY2010,Kharitonov2012}. We note that there are fewer terms here compared to the general results because the single-particle eigenstates have only one nonzero component. The valley anisotropic terms only penalize two electrons that coincide in space. In other words, their valley indices make a difference if the distance between them is comparable to the lattice constant. It is helpful to express the Hamiltonian using the Haldane pseudopotentials $\mathcal{P}_{ij}(\rm{RAM})$~\cite{Haldane1983-3}, which characterize the energy cost associated with putting two electrons in definite relative angular momentum (RAM) states. The Coulomb potential is decomposed to a summation of pseudopotentials in all possible RAM channels, but the valley anisotropic terms only generate the pseudopotential with zero RAM~\cite{Sodemann2014}. The Pauli principle forbids two electrons with the same spin and valley indices to have zero RAM, so the valley anisotropic terms only penalize two electrons with different spin and/or valley indices.

A full description of the zeroth LL physics is quite involved, and the model should be simplified based on experimental evidence to make it numerically tractable~\cite{Zibrov2018}. To begin with, the $\nu=\pm{1/2}$ states are related to each other by particle-hole symmetry within the zeroth LL, so we shall only consider the $\nu=-1/2$ state. It is beneficial to inspect the $\nu=-1$ and $\nu=0$ states. The $\nu=-1/2$ state can be obtained by adding electrons to the $\nu=-1$ state toward the $\nu=0$ state. The spin and valley indices of these electrons should be chosen in such a way that the $\nu=-1$ and $\nu=0$ states are connected. The $\nu=-1$ state is relatively straightforward as the Hamiltonian almost certainly favors a state with complete spin and valley polarizations. In other words, the electrons occupy the same sublevel of the zeroth LL to produce a ferromagnetic state that is expected to have the lowest energy. The nature of the $\nu=0$ state is much more complicated. The proposed candidates include a spin polarized ferromagnetic state, a valley polarized charge density wave state, a canted antiferromagnetic state with broken spin symmetry, and a density wave with partial valley polarization~\cite{Herbut2007,NomuraK2009,WuFC2014,LeeJH2015}. It is quite difficult to distinguish between them in experiments. One piece of useful information is that the $\nu=-1/2$ state should be spin polarized since applying an in-plane magnetic field does not change it~\cite{Zibrov2018}. In contrast, there is no simple way to directly probe the valley index. The physical picture is transparent if spin or valley canting is absent~\cite{Zibrov2018}. When electrons are added on top of the $\nu=-1$ state, they need to populate the same valley to generate a charge density wave but the opposite valley to generate an antiferromagnet. In the transition regime between the charge density wave and the antiferromagnet, the electrons may have equal probability of occupying both valleys. This suggests that we can try to search for the $\nu=-1/2$ FQH state in a two-component system with Coulomb potential modified by the lattice scale valley anisotropic terms. In second quantized notation, the Hamiltonian is
\begin{eqnarray}
H &=& \frac{1}{2} \sum_{\alpha,\beta=\pm}\sum_{\{m_{i}\}} V_{m_{1}m_{2}m_{3}m_{4}} C^{\dag}_{\alpha,m_{1}} C^{\dag}_{\beta,m_{2}} C_{\beta,m_{4}} C_{\alpha,m_{3}} + \frac{F_{0}}{2} \sum_{\{m_{i}\}} W_{m_{1}m_{2}m_{3}m_{4}} C^{\dag}_{+,m_{1}} C^{\dag}_{-,m_{2}} C_{-,m_{4}} C_{+,m_{3}},
\end{eqnarray}
where $C^{\dag}_{\pm,m}$ creates an electron in the $K^{\pm}$ valley with index $m$, the coefficients $V_{m_{1}m_{2}m_{3}m_{4}}$ are derived from the Coulomb potential, and $W_{m_{1}m_{2}m_{3}m_{4}}$ are derived from the zeroth Haldane pseudopotential~\cite{Haldane1983-3}. The energy is measured in units of $e^{2}/(\varepsilon\ell_{B})$ with dielectric constant $\varepsilon$ and magnetic length $\ell_{B}=\sqrt{{\hbar}c/(eB)}$. The factor $F_{0}$ is determined by microscopic physics in a complicated manner, but it will be treated as a phenomenological tunable parameter in our discussion. The eigenstates of this Hamiltonian can be computed using exact diagonalization (ED) and density matrix renormalization group (DMRG). ED is quite straightforward as it employs sparse matrix diagonalization methods to find a few low-lying eigenvalues and the associated eigenstates. DMRG is a variational method that searches for the optimal matrix product representation of the ground state~\cite{White1992,Schollwock2011,Hubig2017,Feiguin2008,ZhaoJZ2011,HuZX2012,Zaletel2013}. The exponential growth of the Hilbert space dimension limits the usage of ED, but DMRG can partially mitigate this problem. The total angular momentum $L^{2}$ and its $z$ component $L_{z}$ are conserved quantities on the sphere. It is obvious that the eigenstates of the two-component system with only Coulomb potential are SU(2) symmetric. One may worry that the valley anisotropic terms break the SU(2) symmetry, but this is not the case because electrons with the same valley index are not affected by the zeroth Haldane pseudopotential.

\section{Results}
\label{result}

\begin{figure}
\centering
\includegraphics[width=0.48\textwidth]{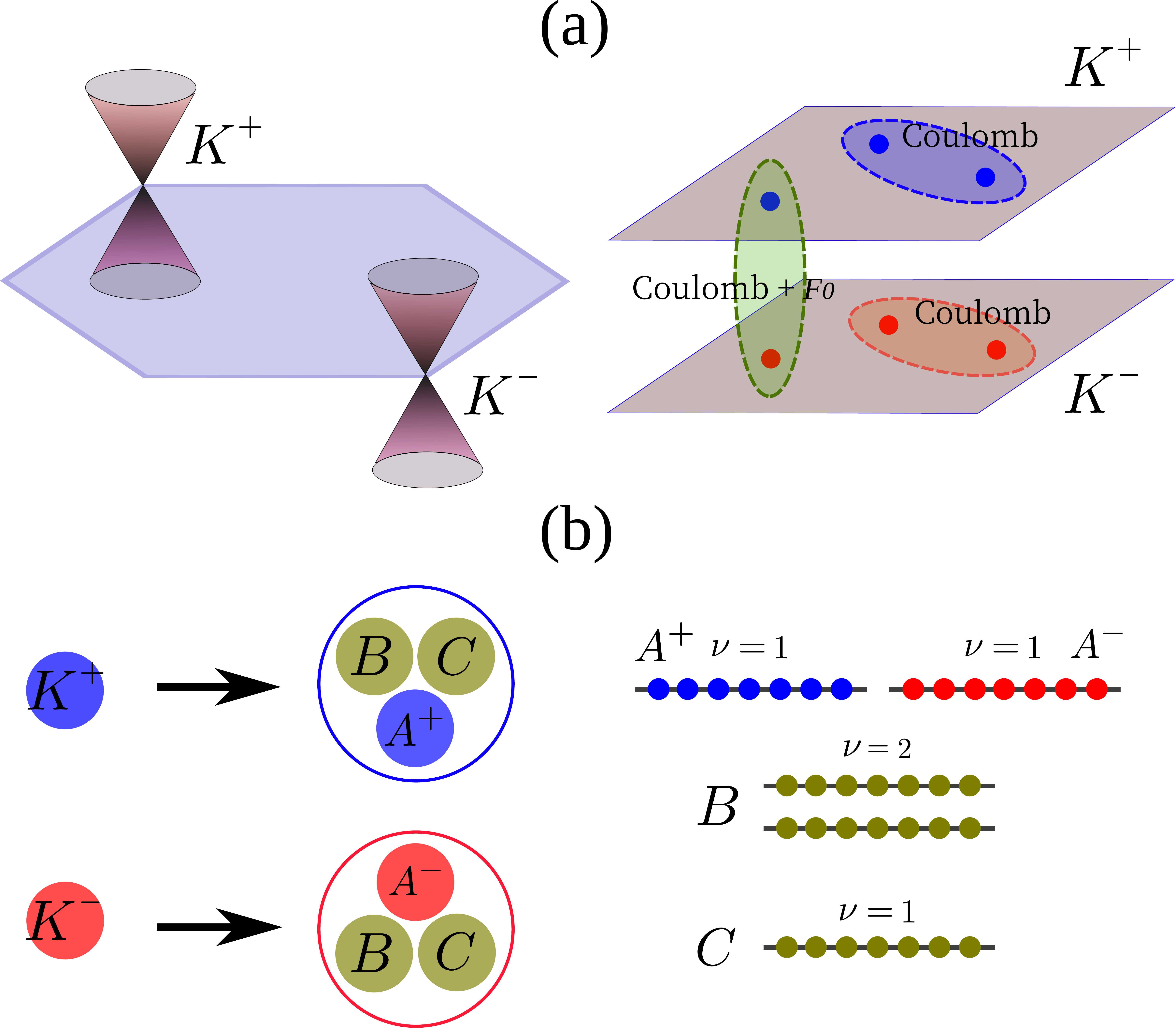}
\caption{Schematics of the valley unpolarized parton state in graphene. (a) In addition to the long-range Coulomb interaction, there is a short-range valley anisotropic interaction between electrons in different valleys. (b) The electrons are decomposed to partons that form their respective IQH states.}
\label{Figure1}
\end{figure}

The total number of electrons is denoted as $N_{e}$ and electrons are distributed equally in the two valleys. To distinguish them, the coordinates of electrons in the $K^{+}$ valley are denoted as $v_{1},v_{2},\ldots,v_{N_{e}/2}$ and those in the $K^{-}$ valley are denoted as $w_{1},w_{2},\ldots,w_{N_{e}/2}$. It is also useful to combine and relabel them as $z_{1} \equiv v_{1} , z_{2} \equiv v_{2} , \ldots , z_{N_{e}/2} \equiv v_{N_{e}/2} , z_{N_{e}/2+1} \equiv w_{1} , z_{N_{e}/2+2} \equiv w_{2} , \ldots , z_{N_{e}} \equiv w_{N_{e}/2}$. The FQH state to be investigated is
\begin{eqnarray}
\Phi \sim \left[ \chi_{1}(\{v\}) \chi_{1}(\{w\}) \right] \chi_{2}(\{z\}) \chi_{1}(\{z\}),
\label{WaveFuncParton}
\end{eqnarray}
which occurs on the sphere at the monopole flux $N_{\phi}=2N_{e}-4$~\cite{Jain1989-2}. The number $4$ is called the shift and the ratio $N_{e}/N_{\phi}$ is not exactly $1/2$ in finite size systems. This wave function can be interpreted using the parton theory as illustrated in Fig.~\ref{Figure1}. One electron is decomposed to three types of partons $A,B,C$ that carry charges $e/4$, $e/4$, $e/2$, respectively. The $A$ type partons inherit the valley index of the electrons while the $B,C$ type partons do not carry such an index. In other words, the electron creation operator is rewritten as
\begin{eqnarray}
\left[ \begin{array}{cc}
\widehat{\psi}^{\dag}_{K^{+}}(\mathbf{r}) \\ 
\widehat{\psi}^{\dag}_{K^{-}}(\mathbf{r})
\end{array} \right] = 
\left[ \begin{array}{cc}
\widehat{\psi}^{\dag}_{A,K^{+}}(\mathbf{r}) \\
\widehat{\psi}^{\dag}_{A,K^{-}}(\mathbf{r}) \\
\end{array} \right] \; \widehat{\psi}^{\dag}_{B}(\mathbf{r}) \; \widehat{\psi}^{\dag}_{C}(\mathbf{r}).
\end{eqnarray}
The two components of the $A$ type partons form the $\nu=1$ IQH states $\chi_{1}(\{v\})$ and $\chi_{1}(\{w\})$, the $B$ type partons form the $\nu=2$ IQH state $\chi_{2}(\{z\})$, and the $C$ type partons form the $\nu=1$ IQH state $\chi_{1}(\{z\})$. The elementary quasihole and quasiparticle of the state should carry charge $e/4$. To see this, one may increase the magnetic flux by one unit on top of the ground state. Because the state has filling factor $1/2$ and is incompressible, this operation creates a localized object with charge $e/2$. In the parton framework, one quasihole was created in both the $\chi_{1}(\{v\})$ and $\chi_{1}(\{w\})$ factors or two quasiholes were created in the $\chi_{2}(\{z\})$ factor. The total charge of two quasiholes is $e/2$ so each one has $e/4$.

\begin{figure}
\centering
\includegraphics[width=0.48\textwidth]{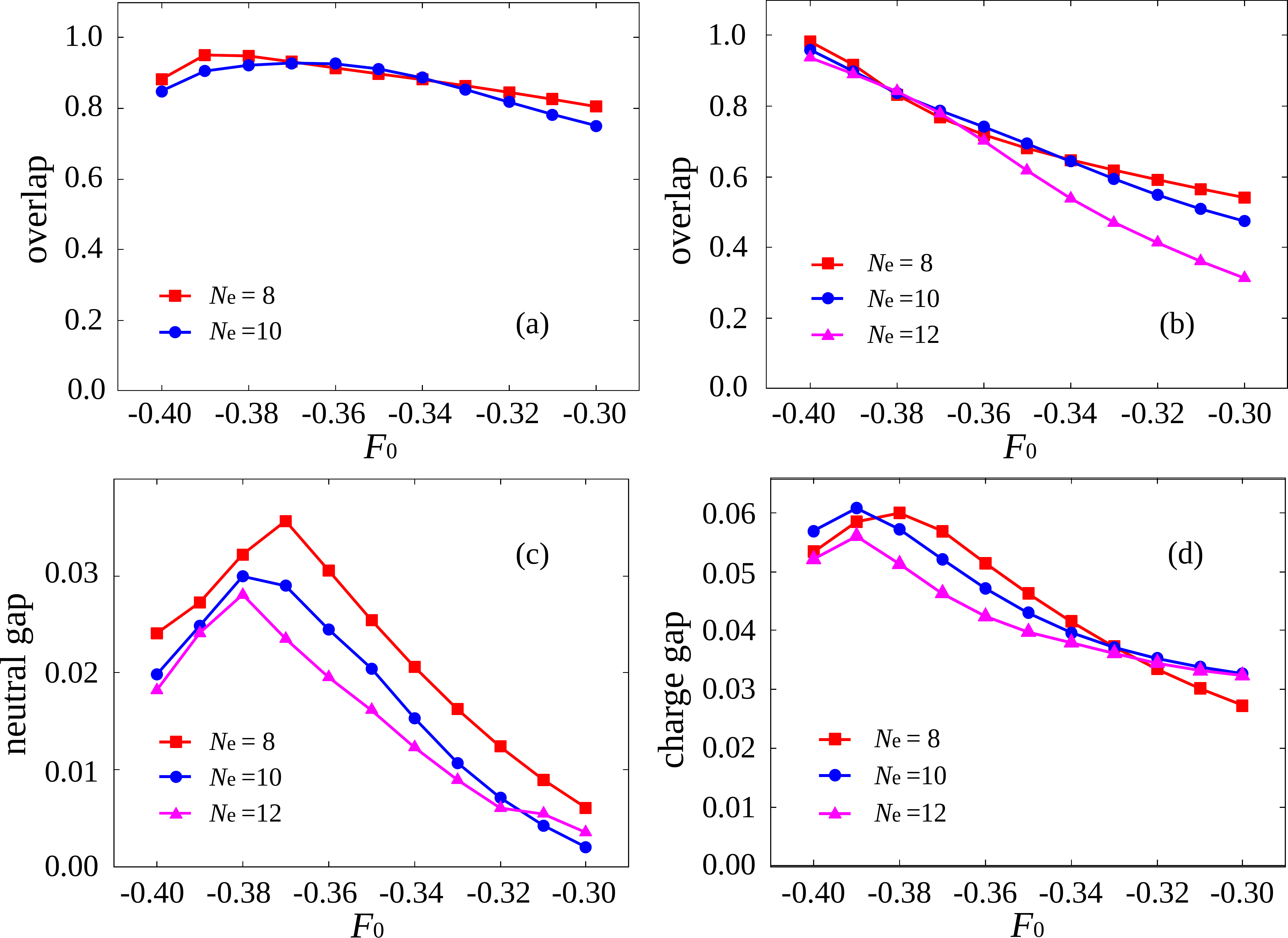}
\caption{Exact diagonalization results for the cases with $N_{e}=8,10,12$ and $F_{0}\in[-0.40,-0.30]$. (a) The overlap between $\Phi_{1}$ in Eq.~(\ref{WaveFuncProj1}) and the exact ground state. (b) The overlap between $\Phi_{2}$ in Eq.~(\ref{WaveFuncProj2}) and the exact ground state. (c) The neutral gap $\Delta_{n}$. (d) The charge gap $\Delta_{c}$.}
\label{Figure2}
\end{figure}

The many-body state Eq.~(\ref{WaveFuncParton}) is not completely in the lowest LL, to which it should be projected in numerical calculations. This is why we have used a $\sim$ sign instead of an equal sign because there are different approaches to the projection~\cite{Dev1992,WuXG1993,Jain1997}. We have considered two different projection methods given by
\begin{eqnarray}
\Phi_{1} = \mathcal{P}_{\rm LLL} \left\{ \left[ \chi_{1}(\{v\}) \chi_{1}(\{w\}) \right] \chi_{2}(\{z\}) \chi_{1}(\{z\}) \right\}
\label{WaveFuncProj1}
\end{eqnarray}
and
\begin{eqnarray}
\Phi_{2} = \left[ \chi_{1}(\{v\}) \chi_{1}(\{w\}) \right] \mathcal{P}_{\rm LLL} \left[ \chi_{2}(\{z\}) \chi_{1}(\{z\}) \right].
\label{WaveFuncProj2}
\end{eqnarray}
The first one means that we multiply the three IQH states of partons, expand it in the Fock space basis, and drop the terms that are not completely inside the lowest LL. The second one means that we multiply the two IQH states of $B$ and $C$ partons, expand it in the Fock space basis, drop the terms that are not completely inside the lowest LL, and then multiply it with the IQH state of $A$ partons. These calculations are very time-consuming so we have only reached $N_{e}=10$ for $\Phi_{1}$ and $N_{e}=12$ for $\Phi_{2}$. One can claim with certain confidence that both wave functions capture identical topological properties, but it is not a priori clear which one has better overlap with exact eigenstates of a specific Hamiltonian. In fact, it is possible that two wave functions in the same universality class have negligible overlap. The overlap between $\Phi_{1}$ ($\Phi_{2}$) and the exact eigenstates for a range of $F_{0}$ are presented in Fig.~\ref{Figure2} (a) [Fig.~\ref{Figure2} (b)]. It turns out that either wave function provides a reasonably good approximation at sufficiently negative $F_{0}$. The maximal value achieved by $\Phi_{1}$ is somewhat lower than that of $\Phi_{2}$, but the former one performs better in a wider range of $F_{0}$. An important difference is that the overlap of $\Phi_{2}$ increases monotonically as $F_{0}$ decreases but that of $\Phi_{1}$ displays a weak peak around $F_{0} \approx -0.38$. There is actually a third possibility
\begin{eqnarray}
\Phi_{3} = \chi_{1}(\{z\}) \mathcal{P}_{\rm LLL} \left\{ \chi_{2}(\{z\}) \left[ \chi_{1}(\{v\}) \chi_{1}(\{w\}) \right] \right\},
\end{eqnarray}
but we find that it does not have appreciable overlaps with exact eigenstates.

\begin{figure}
\centering
\includegraphics[width=0.35\textwidth]{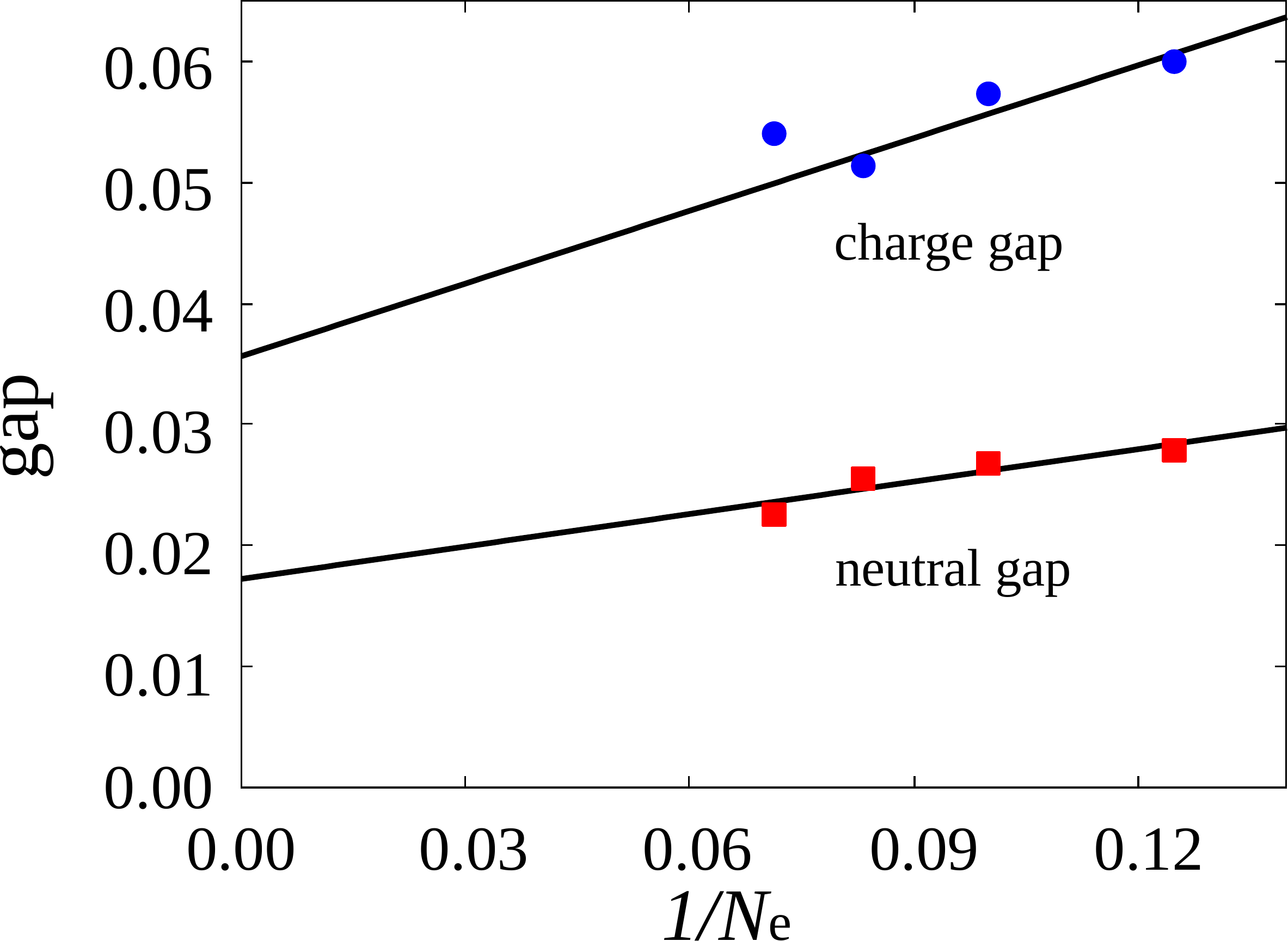}
\caption{Energy gaps $\Delta_{n}$ and $\Delta_{c}$ for the cases with $N_{e}=8,10,12,14$ at $F_{0}=-0.38$. The lines are linear fits of the data points.}
\label{Figure3}
\end{figure}

To establish incompressibility of the state, it is necessary to study the energy gaps of the system. For this purpose, the energy scale should be chosen properly. We have mentioned that the energy is expressed in units of $e^{2}/(\varepsilon\ell_{B})$, but $\ell_{B}$ varies with the system size on the sphere. This problem can be remedied if we rescale the energy eigenvalues using the magnetic length in the thermodynamic limit~\cite{Fano1986,Ambrumenil1989,Morf2002}. The neutral gap
\begin{eqnarray}
\Delta_{n} = E_{1}(N_{e},2N_{e}-4) - E_{0}(N_{e},2N_{e}-4)
\end{eqnarray}
is shown in Fig.~\ref{Figure2} (c). It measures the energy difference between the first excited state, which belongs to a branch of collective magnetoroton mode~\cite{Girvin1986,HeS1994,Platzman1994,ParkK2000-1}, and the ground state. One can see a peak around $F_{0} \approx -0.38$ in Fig.~\ref{Figure2} (c), which basically coincides with the relatively weak peak in Fig.~\ref{Figure2} (a). This implies that $\Phi_{1}$ is a more reliable characterization of the exact ground state. The charge gap $\Delta_{c}$ is shown in Fig.~\ref{Figure2} (d). It is the energy cost associated with a well-separated quasiparticle-quasihole pair, which is closely related to the transport gap measured in experiments~\cite{Fano1986,Ambrumenil1989}. The appearance of a peak around $F_{0} \approx -0.39$ in Fig.~\ref{Figure2} (d) again implies that $\Phi_{1}$ is a better representation of the exact ground state. The neutral gap and charge gap have also been computed for $N_{e}=14$ at $F_{0}=-0.38$ using DMRG. As shown in Fig.~\ref{Figure3}, the gaps do not exhibit very good linear scaling versus $1/N_{e}$, but it is quite plausible that they saturate to finite values in the thermodynamic limit.

The parton interpretation also helps us to understand the edge physics~\cite{WenXG1991-2}. It is helpful to recall some results in the hydrodynamic approach to quantum Hall edge physics. The edge wave is described by a one-dimensional density function $\rho(x)$ with the equation $\partial_{t} \rho + v \partial_{x} \rho = 0$. After canonical quantization, we obtain the $U(1)$ Kac-Moody algebra
\begin{eqnarray}
\left[ \rho(k),\rho(k') \right] = \nu \frac{k}{2\pi} \delta_{k+k'}
\end{eqnarray}
for the momentum space density function $\rho(k)$. In the parton construction, a subscript is introduced to distinguish the different types of partons, and the commutation relation becomes
\begin{eqnarray}
\left[ \rho_{\lambda}(k), \rho_{\mu}(k') \right] = \frac{k}{2\pi} \delta_{\lambda\mu} \delta_{k+k'}
\end{eqnarray}
The partons in Eq.~(\ref{WaveFuncParton}) have five edge modes: $\rho_{1}$ from $\chi_{1}(\{v\})$, $\rho_{2}$ from $\chi_{1}(\{w\})$, $\rho_{3,4}$ from $\chi_{2}(\{z\})$, and $\rho_{5}$ from $\chi_{1}(\{z\})$. These modes are not independent and they should be subjected to the constraint that relative density oscillations vanish~\cite{WenXG1991-2}. To this end, we introduce an operator $\widetilde{\rho}_{C} = C_{1} \left( \rho_{1} + \rho_{2} \right) + C_{2}(\rho_{3}+\rho_{4}) + C_{3} \rho_{5}$. If $\sum^{3}_{\alpha=1} C_{\alpha}=0$, $\widetilde{\rho}_{C}$ must commute with any physical operator because the fluctuations associated with $\widetilde{\rho}_{C}$ are unphysical. One can check that the edge density operators 
\begin{eqnarray}
j_{0} = \sqrt{\frac{1}{2}} \left[ \frac{1}{2} \left( \rho_{1}+\rho_{2} \right) + \frac{1}{2} \left( \rho_{3} + \rho_{4} \right) + \rho_{5} \right], \quad j_{1} = \sqrt{\frac{1}{2}} \left( \rho_{1} - \rho_{2} \right), \quad j_{2} = \sqrt{\frac{1}{2}} \left( \rho_{3} - \rho_{4} \right),
\end{eqnarray}
commute with $\widetilde{\rho}_{C}$, so the physical edge excitations have three branches. This means that the thermal Hall conductance of this state should be $3$ in units of $\pi^{2}k^{2}_{B}T/(3h)$. It should be emphasized that our exploration of the edge physics is preliminary and things could be much more complicated in actual samples. If the confinement potential due to positive charges is not sharp enough in a system, its edge may be reconstructed such that additional modes are generated~\cite{MacDonald1993,Chamon1994,WanX2002,Joglekar2003,YangK2003,WangJH2013,Sabo2017}. In the simplest scenario, edge reconstruction does {\em not} change the thermal Hall conductance. The presence of strong disorder further complicates things as electric and thermal transport properties could become sample-size dependent~\cite{Kane1994,Moore1998,Protopopov2017}. It is also desirable to search for other probes that can be used in combination with thermal Hall conductance. Shot noise has been proposed as a good indicator for the $5/2$ state~\cite{ParkJH2020} and may also be useful in the current context. 

Besides the parton state Eq.~(\ref{WaveFuncParton}), there are two other celebrated two-component FQH states at filling factor $1/2$. The first one is the Halperin 331 state~\cite{Halperin1983} 
\begin{eqnarray}
\Phi_{331} = \prod_{j<k} (v_{j}-v_{k})^{3} \prod_{j<k} (w_{j}-w_{k})^{3} \prod_{j,k} (v_{j}-w_{k})
\label{WaveFunc331}
\end{eqnarray}
with parent Hamiltonian
\begin{eqnarray}
\sum_{i,j \in K^{+}} P_{ij}(1) + \sum_{i,j \in K^{-}} P_{ij}(1) + \sum_{i \in K^{+}, j \in K^{-}} P_{ij}(0).
\label{Hamilton331}
\end{eqnarray}
The second one is the Haldane-Rezayi state~\cite{Haldane1988-3}
\begin{eqnarray}
\Phi_{\rm HR} = {\rm Det} \left[ \frac{1}{(v_{j}-w_{k})^{2}} \right] \prod_{j<k} (z_{j}-z_{k})^{2}
\label{WaveFuncHR}
\end{eqnarray}
with parent Hamiltonian
\begin{eqnarray}
\sum_{i,j \in K^{\pm}} P_{ij}(1) .
\label{HamiltonHR}
\end{eqnarray}
We believe that they are not reasonable candidates for the $\nu={\pm}1/2$ states in graphene. One important reason that disfavors $\Phi_{331}$ is that it does not have the same SU(2) symmetry as the exact ground state. In previous studies, it has been found that $\Phi_{331}$ can be realized in bilayer systems when the inter-species interaction is reduced to some extent compared to the intra-species interaction~\cite{HeS1993}. This scenario is also different from the graphene system because only the zeroth Haldane pseudopotential is reduced. The Haldane-Rezayi state can be constructed using non-unitary conformal field theory and it is most likely gapless in the thermodynamic limit~\cite{Gurarie1997,Read2000,Hermanns2011,Crepel2019}. The shift of $\Phi_{331}$ is $3$ and that of $\Phi_{\rm HR}$ is $4$, so the former cannot be compared directly with the parton state but the latter can. The two trial states with $N_{e}=12$ have been generated numerically. The overlaps with the exact ground state are not impressive: the maximal value is $0.2339$ for $\Phi_{331}$ (at $F_{0}=-0.30$) and $0.4376$ for $\Phi_{\rm HR}$ (at $F_{0}=-0.40$). Thermal Hall conductance of the 331 state is $2$ in units of $\pi^{2}k^{2}_{B}T/(3h)$, which can be used to distinguish it from the parton state in experiments. If the Haldane-Rezayi state is really gapless, there would be no quantized thermal Hall conductance.

\vspace{1em}

\section{Conclusions}
\label{conclude}

In conclusion, we propose that the $\nu={\pm}1/2$ states observed by Zibrov et al.~\cite{Zibrov2018} can be understood using the two-component parton wave function Eq.~(\ref{WaveFuncParton}). This claim is supported by numerical calculations in a model of graphene with valley anisotropic interactions. The importance of valley anisotropic interactions calls for more in-depth studies. The parton framework can generate many non-Abelian FQH states and some of them may be experimentally relevant. It is natural to ask how to construct multi-component non-Abelian FQH states using the parton theory and search for appropriate conditions to realize them. We hope that this paper will motivate further investigations of parton FQH states in various platforms. 

\section*{Acknowledgements}

Exact diagonalization is performed using the DiagHam package, for which we are grateful to the authors. This work was supproted by the NNSF of China under grant No. 12174130.

\bibliography{../ReferCollect}

\end{document}